\documentclass[twocolumn,showpacs,amsmath,nofootinbib]{revtex4-2}
\pdfoutput=1
\usepackage{float} 
\usepackage{amsmath}
\usepackage{amsfonts}
\usepackage{amssymb}
\usepackage{amsthm}
\usepackage{graphicx}
\usepackage[colorlinks,citecolor=blue]{hyperref}
\usepackage{color}
\usepackage{flafter}
\usepackage[normalem]{ulem}

\usepackage[toc,page]{appendix}
\usepackage{makecell}
\usepackage{slashed}
\usepackage{tikz-cd}
\usepackage{makecell}
\usepackage{subcaption}
\usepackage{graphicx}
\usepackage{xr}
\usepackage[font=small,labelfont=bf,
justification=justified,
format=plain]{caption}

\allowdisplaybreaks

\definecolor{OliveGreen}{rgb}{0,0.6,0}
\definecolor{Orange}{rgb}{1.00, 0.65, 0}
\definecolor{Grey}{rgb}{0.43, 0.5, 0.5}

\newcommand{\Fig}[1]{Fig.~\ref{#1}}
\newcommand{\Eq}[1]{Eq.(\ref{#1})}
\newcommand{\nn}{\nonumber\\}

\newcommand{\be}{\begin{eqnarray}}
\newcommand{\ee}{\end{eqnarray}}
\newcommand{\bpm}{\begin{pmatrix}}
\newcommand{\epm}{\end{pmatrix}}

\newcommand{\p}{\partial}

\newcommand{\tr}{{\rm tr}}

\newcommand{\ra}{\rightarrow}

\renewcommand{\v}[1]{{\boldsymbol{#1}}}

\newcommand{\e}{\epsilon}
\newcommand{\s}{\sigma}
\renewcommand{\t}{\tau}
\newcommand{\g}{\gamma}
\newcommand{\G}{\Gamma}

\newcommand{\comment}[1]{}

\begin{document}
\title{Non-abelian bosonization of fermion symmetry-protected topological states}
\author{Yen-Ta Huang$^{1}$}
\author{Dung-Hai Lee$^{1,2}$}\email{Corresponding author: dunghai@berkeley.edu}

\affiliation{
$^1$ Materials Sciences Division, Lawrence Berkeley National Laboratory, Berkeley, CA 94720, USA.\\
$^2$ Department of Physics, University of California, Berkeley, CA 94720, USA.\\
}

\begin{abstract}
Applying the results of Ref.\cite{Huang2021}, we carry out the non-abelian bosonization for a class of free fermion symmetry-protected topological states (SPTs). The resulting boson theories are non-linear sigma models with topological $\theta$ term, thus unifying the field theory description of the (free) fermionic and bosonic SPTs. Tuning $\theta$ from $0$ to $2\pi$ triggers the topological trivial to non-trivial phase transitions. In addition, applying the same idea to the symmetry-protected critical spin liquids, we obtain the NL$\s$ models for spin SPTs living in one spatial dimension higher.
\end{abstract}
\maketitle

{\bf{Introduction}}
Symmetry-protected topological states (SPTs) are a new frontier in condensed matter physics. Broadly speaking SPTs can be divided into two types: the bosonic ones \cite{Chen2012} and the fermionic ones \cite{Schnyder2008,Kitaev2009,Hasan2010}. A well-known (and experimentally realized) bosonic SPT is the spin 1 chain \cite{Haldane1983}. As long as the protection symmetry, e.g., SO(3), is unbroken, the boundary of the spin 1 chain is spin 1/2s. 
The fermionic SPTs are much wider experimentally realized. Like the bosonic SPT, as long as the protection symmetry is unbroken, the boundary always harbors gapless fermion modes. However, despite the similarity, the (continuum) quantum field theories used to describe bosonic and fermionic SPTs are very different. Non-linear sigma (NL$\s$) models with topological terms are often used for bosonic SPTs. In contrast, fermion SPTs are described in terms of massive relativistic fermion theories. The purpose of this paper is to present a unification of the field theory descriptions by bosonizing the fermion SPTs.

In Ref.\cite{Huang2021}, the present authors carried out non-abelian bosonization for {\it massless} relativistic fermions 
in spatial dimensions $d=1,2,3$ (in the rest of the paper $d$ and $D$ denote the space and space-time dimensions, respectively). These massless fermions are protected by a {\it maximum set of emergent symmetries}\footnote{Maximum in the sense that all symmetries leaving the massless Dirac (Majorana) action invariant are included.}. They can be the low energy effective theory of lattice problems or be realized at the boundary of topological insulators/superconductors. Because of the maximum protection symmetries, all such SPTs in are $\mathbb{Z}$-classified \cite{Huang2021}. In this paper, we bosonize these fermionic SPTs. The results are non-linear sigma (NL$\s$) models with topological $\theta$ term. Such topological term is equal to $2\pi i$ times the wrapping number associated with the mapping from the space-time to the order parameter manifold (OPM). Their boundary gapless fermions, on the other hand, are bosonized by NL$\s$ models with level-1 Wess-Zumino-Witten (WZW) terms \cite{Huang2021}.

Making an analogy with the spin chains, the topologically trivial fermion SPT is analogous to the spin 0 chain, and the topologically non-trivial fermion SPT is analogous to the spin 1 chain. The difference is that the order parameter manifold (OPM) for the bosonized fermionic SPT is more complicated than that of the spin chains (whose OPM is $S^2$). Of course, in spatial dimensions greater than one the boundary is no longer a set of points. Similar to spin chains, under the open boundary condition, the bosonized boundary action are NL$\s$ models with WZW terms. Here the WZW term manifests the Berry phase of massless fermions instead of spin 1/2s. 

In addition to the above, we also outline the derivation for the bulk spin SPTs whose boundaries are symmetry-protected critical spin liquids. The best known example of critical spin liquids 
include the $d=1$ NL$\s$ model with $S^3$ as the OPM, and $d=2$ NL$\s$ models with $S^4$ as OPM (describing the deconfined quantum critical point \cite{Senthil2004,Tanaka2005} ). They each have a level-1 WZW term. 
If we view each of these critical spin liquids as being protected by emergent symmetries, namely, $O(4), O(5)$ for $d=1$ and $2$, respectively, they too can be viewed as the boundary of spin SPTs with topological $\theta$ terms\cite{Chen2011,Chen2013}. 

In the rest of the paper, we shall focus on fermion SPTs which have gapless {\it non-chiral} boundary modes. Besides the flavor symmetry, these fermions have the $Q,T,C$ symmetries\footnote{$Q,T,C$ stands for charge conservation, time reversal, and charge conjugation.} considered in the ``ten-fold way'' classification \cite{Schnyder2008,Kitaev2009}. Moreover, we shall restrict ourselves to the theories where the fermion field has the smallest number of components necessary to represent the $Q,T,C$ and the flavor symmetries. These fermion theories serve as the ``generator'' for the ``stacking operation'' in the classification of SPTs. In our case since the topological classification is $\mathbb{Z}$, it plays the role of the number ``1'' under addition in the group $\mathbb{Z}$. 

The Hamiltonian of the fermion SPTs under consideration has the following form 
\be 
H=\int d^d x~\psi^{\dagger}(\v x)\left[-i\sum_{i=1}^d \G_i\p_i-m_0 M_B\right]\psi(\v x).\label{h}\ee
Here $\psi$ is the real (Majorana) or complex fermion field depending on whether we are talking about topological insulators or superconductors. Including $n$ flavors, $\psi$ has $2n,4n,4n$ components in $d=1,2,3$. In table SM-I 
of the supplementary material we list the $\G_i$, $M_B$ and the full protection symmetries of these SPTs. In this table the ``complex'' and the ``real'' classes refer to topological insulators and superconductors. We shall also focus on $n\ge n_c$ so that after bosonization the boundary WZW term exists ($n_c=4,3,6$ for $d=0,1,2$ in the real class, and $n_c=2,2,4$ for $d=0,1,2$ in the complex class). 

\begin{figure}[h]
\includegraphics[scale=0.3]{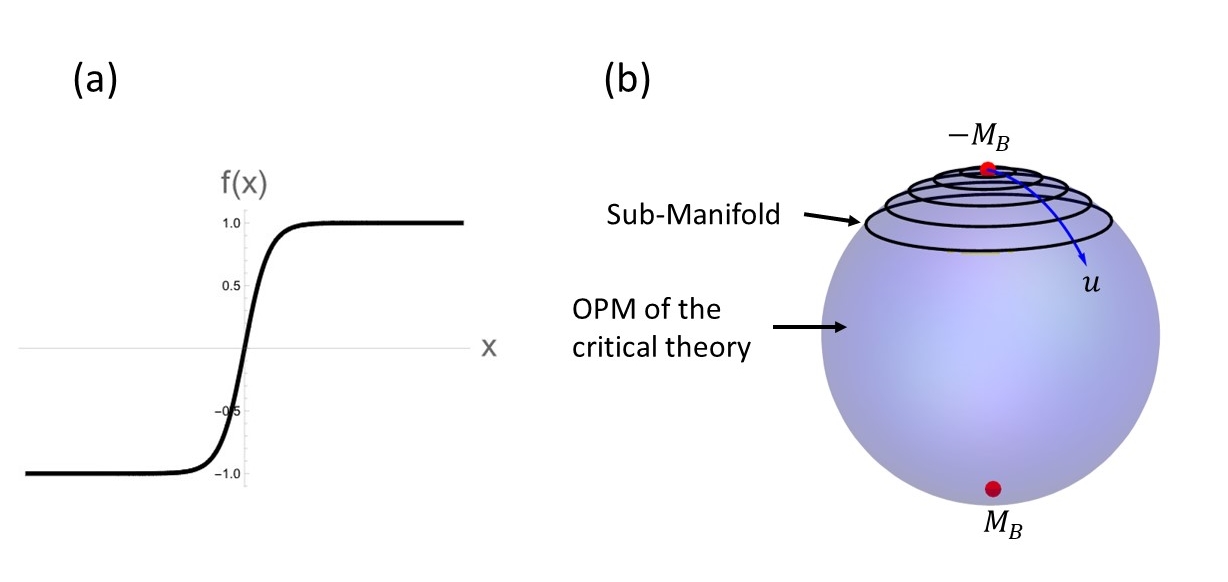}
\caption{(a) The mass profile of a domain wall. (b) Schematic representation of the OPM of the critical theory (the light blue sphere), and the one-parameter family of sub-manifolds in OPM (black circles) were used to derive the $\theta$ term from the WZW term. The red points represent $\mp M_B$ and the blue arrowed arc indicates the direction of the increasing parameter ($u$).} 
\label{combined}
\end{figure}

{\bf{The boundary gapless fermion modes}}
The boundary of the fermion SPTs in Table SM-I 
can be modeled by a domain wall where $m_0$ changes sign. 
The Hamiltonian is given by \Eq{h} except $m_0$ is replaced with $m_0 f(x_d)$, where $x_d$ is, say, the last spatial coordinate. The function $f(x)$ is shown in \Fig{combined}(a).

On the domain wall 
the gapless fermion modes are governed by \cite{Tsui2019} \be
H_{\rm dw}=\int d^{d-1}x~\chi^{\dagger}(\v x)\left[-i\sum_{i=1}^{d-1}\g_i\p_i\right]\chi(\v x).
\label{gapless}
\ee
In the above 
$\g_i=\mathcal{P}\G_i\mathcal{P}$, $\chi=\mathcal{P}\psi,$ and the boundary symmetry generators are equal to $\mathcal{P}~{\rm (bulk~symmetry~generators)~}\mathcal{P}$, where the projection operator is given by $$\mathcal{P}=\left({i\G_d M_B+I}\right)/2.$$ Due to the projection, the number of components in $\chi$ is half of that in $\psi$.
The boundary $\g$ matrices and symmetry generators are summarized in Table SM-II 
of the supplementary material.

{\bf{Bosonized theory for the boundary of SPTs}}
The boundary gapless fermions in Table SM-II 
are exactly the subjects of non-abelian bosonization in Ref.\cite{Huang2021}. The results are NL$\s$ models with level-1 WZW terms.
The action and the OPM of such NL$\s$ models are summarized in Table \ref{NLS} of the main text.
\begin{table*}
\small
\renewcommand{\arraystretch}{2}
\begin{tabular}{ |c|c|c| }
\hline
$D=0+1$ & Real class & Complex class \\
\hline
NL$\s$ model action & ${1\over 2g} \int_{S^1} d\t \, \tr\left[ \partial_{\mu} Q^{\mathbb{R}} \partial^{\mu}Q^{\mathbb{R}} \right]-W_{\rm WZW}$ &$
{1\over 2g} \int_{S^1} d \t \, \tr\left[ \partial_{\mu} Q^{\mathbb{C}} \partial^{\mu}Q^{\mathbb{C}} \right]-W_{\rm WZW}$ \\
\hline
$W_{\rm WZW}$ &$\frac{2 \pi}{32 \pi} \int\limits_{D^2} \tr \Big[ \tilde{Q}^{\mathbb{R}} \, (d \tilde{Q}^{\mathbb{R}})^2 \Big]$ &$\frac{ 2 \pi }{16 \pi} \int\limits_{D^2} \, \tr \Big[\tilde{Q}^{\mathbb{C}} \,\left( d \tilde{Q}^{\mathbb{C}} \right)^2 \Big]$\\
\hline
OPM &$Q^{\mathbb{R}}\in \frac{O(n)}{U(n/2)}$&$Q^{\mathbb{C}}\in \frac{U(n)}{U(n/2)\times U(n/2)}$\\
\hline 
\Xhline{3\arrayrulewidth}
$D=1+1$ & Real class & Complex class \\
\hline
NL$\s$ model action &$\frac{1}{16\pi} \int_{S^2} d^2x \, \tr\left[ \partial_\mu Q^{\mathbb{R}} \partial^\mu (Q^{\mathbb{R}})^T \right] -W_{\rm WZW}$&
$
\frac{1}{8\pi} \int_{S^2} d^2x \, \tr\left[ \partial_\mu Q^{\mathbb{C}} \partial^\mu Q^{\mathbb{C}\dagger} \right]-W_{\rm WZW}$\\
\hline
$W_{\rm WZW}$ &$\frac{2 \pi i}{48\pi^2} \int\limits_{D^3} \, \tr \Big[ \left[(\tilde{Q}^{\mathbb{R}})^T d \tilde{Q}^{\mathbb{R}}\right]^3 \Big]$ &$\frac{2\pi i}{24\pi^2} \int\limits_{D^3} \, \tr \Big[ \left( \tilde{Q}^{\mathbb{C}\dagger} d \tilde{Q}^{\mathbb{C}} \right)^3 \Big]$\\
\hline
OPM &$Q^{\mathbb{R}}\in O(n)$&$Q^{\mathbb{C}}\in U(n)$\\
\Xhline{3\arrayrulewidth}
$D=2+1$ & Real class & Complex class \\
\hline
NL$\s$ model action &${1\over 2g} \int_{S^3} d^3 x \, \tr\left[ \partial_{\mu} Q^{\mathbb{R}} \partial^{\mu}Q^{\mathbb{R}} \right]-W_{\rm WZW}$&
$
{1\over 2g} \int_{S^3} d^3 x \, \tr\left[ \partial_{\mu} Q^{\mathbb{C}} \partial^{\mu}Q^{\mathbb{C}} \right]-W_{\rm WZW}$\\
\hline
$W_{\rm WZW}$ &$\frac{2 \pi i}{512 \pi^2} \int\limits_{D^4} \tr \Big[ \tilde{Q}^{\mathbb{R}} \, (d \tilde{Q}^{\mathbb{R}})^4 \Big]$ &$\frac{ 2 \pi i }{256 \pi^2} \int\limits_{D^4} \, \tr \Big[\tilde{Q}^{\mathbb{C}} \,\left( d \tilde{Q}^{\mathbb{C}} \right)^4 \Big]$
\\
\hline
OPM &$Q^{\mathbb{R}}\in \frac{O(n)}{O(n/2) \times O(n/2)}$&$Q^{\mathbb{C}}\in \frac{U(n)}{U(n/2)\times U(n/2)}$\\
\hline 
\end{tabular}
\caption{A summary of the bosonized boundary theory from Ref.\cite{Huang2021}.} 
\label{NLS}
\end{table*}

In Table \ref{NLS} the boundary space-time manifold $\mathcal{M}$ is $S^{D-1}$ and $\mathcal{B}$ is a $D$ dimension ``disk'' such that $\p\mathcal{B}=\mathcal{M}$. The order parameter $Q^{\mathbb{C,R}}$ are matrices in the spaces given in the rows marked ``OPM''.
The WZW term is the Berry's phase difference between a space-time constant $Q^{\mathbb{C}}$ or $Q^{\mathbb{R}}$, and the $Q^{\mathbb{C,R}}(x^\mu)$ in question. 

For example, in the $D=(2+1)$ complex class 
\be
W_{\rm WZW}[\tilde{Q}^{\mathbb{C}}] =\frac{ 2 \pi i }{256 \pi^2} \int \limits_{\mathcal{B}} \, \tr \left[\tilde{Q}^{\mathbb{C}} \,\left( d \tilde{Q}^{\mathbb{C}} \right)^4 \right],
\label{wzw10}\ee
where $Q^{\mathbb{C}}, \tilde{Q}^{\mathbb{C}}$ are matrices in $\frac{U(n)}{U(n/2)\times U(n/2)}$. 
In \Eq{wzw10} $\tilde{Q}^{\mathbb{C}} (\tau,x,y,u)$ is a one-parameter extension of the space-time configuration $Q^{\mathbb{C}}$. At $u=0$, $\tilde{Q}^{\mathbb{C}}(\tau,x,y,0)$ is a space-time independent matrix. At $u=1$ the $\tilde{Q}^{\mathbb{C}}=Q^{\mathbb{C}}(x^\mu)$. Moreover, in \Eq{wzw10} \be &&\int \limits_{\mathcal{B}} \rightarrow\int_0^1du\int_{S^3} d^3x\nn&&\tilde{Q}^{\mathbb{C}} \,\left( d \tilde{Q}^{\mathbb{C}} \right)^4 =~\e^{\mu\nu\rho\lambda}
\tilde{Q}^{\mathbb{C}}\p_\mu\tilde{Q}^{\mathbb{C}}\p_\nu\tilde{Q}^{\mathbb{C}} \p_\rho\tilde{Q}^{\mathbb{C}} \p_\lambda\tilde{Q}^{\mathbb{C}},\nonumber\ee where $\mu,\nu,\rho,\lambda\in {\t,x,y,u}$. 
Physically the WZW term is the Berry phase accumulated during the adiabatic deformation from $\tilde{Q}^{\mathbb{C}}(\tau,x,y,0)$ to $\tilde{Q}^{\mathbb{C}}(\tau,x,y,1)$. It can be shown that $\exp\left(-W_{WZW}\right)$ is independent of the $u<1$ values $\tilde{Q}^{\mathbb{C}}(\tau,x,y,u)$, as long as the coefficient in \Eq{wzw10} is an integer multiple of $\frac{ 2 \pi i }{256 \pi^2}$.\footnote{For the WZW term to be well-defined it requires any space-time configuration of the order parameters can be smoothly deformed into a trivial (space-time constant) configuration. This requires the homotopy group of the space-time to the OPM map to be trivial $$\pi_{D}\left(OPM\right)=0.$$ To ensure $\exp\left(-W_{WZW}\right)$ to be independent of the specific path of interpolation it requires $$\pi_{D+1}\left(OPM\right)=\mathbb{Z}.$$ These requirements are satisfied by the homotopy groups of all the OPM in Table \ref{NLS} of the main text.}

{\bf{The NL$\s$ models with topological $\theta$ terms}}
As discussed in the preceding paragraph, on the boundary $\exp\left(-W_{\rm WZW}\right)$ only depends on the value of 
$\tilde{Q}^{\mathbb{C,R}}(x^\mu,u=1)$. Here $x^\mu\in S^{D-1}$ (the boundary space-time manifold). 
To prove that we consider two different $\tilde{Q}^{\mathbb{C,R}}(x^\mu,u)$, namely, $\tilde{Q}_1^{\mathbb{C,R}}(x^\mu,u)$ and $\tilde{Q}_2^{\mathbb{C,R}}(x^\mu,u)$, with $$\tilde{Q}_1^{\mathbb{C,R}}(x^\mu,u=1)=\tilde{Q}_2^{\mathbb{C,R}}(x^\mu,u=1).$$
Since the WZW term is purely imaginary, the relative phase factor $\exp\left(-W_{\rm WZW}\right)$ associated with $\tilde{Q}_{1,2}^{\mathbb{C,R}}$ is
\be\exp\left(-W_{\rm WZW}[\tilde{Q}_{1}^{\mathbb{C,R}}]+W_{\rm WZW}[\tilde{Q}_{2}^{\mathbb{C,R}}]\right).\label{rpf}\ee
Since the WZW term involves $\p_u$, reversing the integration limit in $u$ changes the sign of the WZW term. Consequently we can regard $W_{\rm WZW}[\tilde{Q}_{1}^{\mathbb{C,R}}]-W_{\rm WZW}[\tilde{Q}_{2}^{\mathbb{C,R}}]$ as the integral from $u=0$ to $u=1$ back to $u=0$ (recall $\tilde{Q}_{1,2}^{\mathbb{C,R}}$ agree at $u=1$). This is the WZW term defined on the closed manifold $S^{D}$. The condition that the $\exp\left(-W_{\rm WZW}\right)$ is well defined (i.e., only depends on the value of $\tilde{Q}^{\mathbb{C,R}}$ at $u=1$) requires \Eq{rpf} be $1$, or $W_{\rm WZW}[\tilde{Q}_{1}^{\mathbb{C,R}}]-W_{\rm WZW}[\tilde{Q}_{2}^{\mathbb{C,R}}]$ is an integer multiple of $2\pi i$. In fact, when defined on the closed manifold $S^D$, the value of $W_{\rm WZW}/2\pi i$ is precisely the wrapping number, of the $S^{D}\ra \text{OPM}$ map. For our case 
$\pi_{D}\left(\text{OPM}\right)=\mathbb{Z}$. 

Thus $W_{\rm WZW}/2\pi i$ is precisely the topological invariant $W$ in Table \ref{NLS2} of the main text. The WZW term in Table \ref{NLS} of the main text can be interpreted as the topological $\theta$ term when the space-time manifold is cut open in the direction coordinated by $u$. In short, the NL$\s$ model with the WZW term is just the boundary action of a bulk NL$\s$ model with the $\theta$ term. This is summarized in Table \ref{NLS2} of the main text.
\begin{table*}
\small
\renewcommand{\arraystretch}{2}
\begin{tabular}{ |c|c|c| }
\hline
$D=1+1$ & Real class & Complex class \\
\hline
NL$\s$ model action & ${1\over 2g} \int_{S^2} d^2 x \, \tr\left[ \partial_{\mu} Q^{\mathbb{R}} \partial^{\mu}Q^{\mathbb{R}} \right]-W_\theta$ &$
{1\over 2g} \int_{S^2} d^2 x\, \tr\left[ \partial_{\mu} Q^{\mathbb{C}} \partial^{\mu}Q^{\mathbb{C}} \right]-W_\theta$\\
\hline
$W_\theta$ &
\thead{$\frac{2\pi}{32 \pi} \int\limits_{S^2} \tr \Big[ Q^{\mathbb{R}} \, (d Q^{\mathbb{R}})^2 \Big]$ }&
\thead{$\frac{ 2\pi}{16 \pi} \int\limits_{S^2} \, \tr \Big[Q^{\mathbb{C}} \,\left( d Q^{\mathbb{C}} \right)^2 \Big]$}\\
\hline
OPM &$Q^{\mathbb{R}}\in \frac{O(n)}{U(n/2)}$&$Q^{\mathbb{C}}\in \frac{U(n)}{U(n/2)\times U(n/2)}$ \\
\Xhline{3\arrayrulewidth}
$D=2+1$ & Real class & Complex class \\
\hline
NL$\s$ model action &$\frac{1}{16\pi} \int_{S^3} d^3x \, \tr\left[ \partial_\mu Q^{\mathbb{R}} \partial^\mu (Q^{\mathbb{R}})^T \right] -W_\theta$&
$
\frac{1}{8\pi} \int_{S^3} d^3x \, \tr\left[ \partial_\mu Q^{\mathbb{C}} \partial^\mu Q^{\mathbb{C}\dagger} \right]-W_\theta$\\
\hline
$W_\theta$ &
$\frac{2\pi i}{48\pi^2} \int\limits_{S^3} \, \tr \Big[ \left((Q^{\mathbb{R}})^T d Q^{\mathbb{R}}\right)^3 \Big]$ &
$\frac{2\pi i}{24\pi^2} \int\limits_{S^3} \, \tr \Big[ \left( Q^{\mathbb{C}\dagger} d Q^{\mathbb{C}} \right)^3 \Big]$\\
\hline
OPM &$Q^{\mathbb{R}}\in O(n)$&$Q^{\mathbb{C}}\in U(n)$\\
\Xhline{3\arrayrulewidth}
$D=3+1$ & Real class & Complex class \\
\hline
NL$\s$ model action &${1\over 2g} \int_{S^4} d^4 x \, \tr\left[ \partial_{\mu} Q^{\mathbb{R}} \partial^{\mu}Q^{\mathbb{R}} \right]-W_\theta$&
$
{1\over 2g} \int_{S^4} d^4 x \, \tr\left[ \partial_{\mu} Q^{\mathbb{C}} \partial^{\mu}Q^{\mathbb{C}} \right]-W_\theta$\\
\hline
$W_\theta$ &$\frac{2\pi i}{512 \pi^2} \int\limits_{S^4} \tr \Big[ Q^{\mathbb{R}} \, (d Q^{\mathbb{R}})^4 \Big]$ &
$\frac{2\pi i}{256 \pi^2} \int\limits_{S^4} \, \tr \Big[Q^{\mathbb{C}} \,\left( d Q^{\mathbb{C}} \right)^4 \Big]$
\\
\hline
OPM &$Q^{\mathbb{R}}\in \frac{O(n)}{O(n/2) \times O(n/2)}$&$Q^{\mathbb{C}}\in \frac{U(n)}{U(n/2)\times U(n/2)}$ \\
\hline 
\end{tabular}
\caption{The bosonized bulk SPT having the NL$\s$ models in Table \ref{NLS} as boundary. } 
\label{NLS2}
\end{table*}

{\bf{An alternative Derivation of the $\theta$ term}}
In the preceding paragraph, we discussed the relation between the $\theta$ term in $D$ and the WZW term in $D-1$ space-time dimensions. Here we show the $\theta$ term can also be derived from the WZW term in the {\it same} $D$ space-time dimensions.

Taking the SPT given in Table SM-I 
and setting $m_0$ in \Eq{h} to zero, we obtain massless fermion theories describing the critical point of phase transitions between SPTs described by $m_0>0$ and $m_0<0$. We can apply the method of Ref.\cite{Huang2021} to bosonize such a critical theory. The results are given in Table SM-III 
of the supplementary material. In \Fig{combined}(b) the OPM of the critical theory is schematically represented by the light blue sphere, and $\pm M_B$ are represented as the two red points.

If the order parameter fluctuates uniformly in the full OPM of the critical theory, the {\it emergent} symmetry is that of the fermion theory at $m_0=0$ \cite{Huang2021}, which is larger than the protection symmetries in Table SM-I
, which are for $m_0\ne 0$. 
In the following, we select a family of sub-manifolds in the OPM of the critical theory. These sub-manifolds are parametrized by $u$ and have the properties that (i) at $u=0$ and $u=1$ the sub-manifold shrinks to $\mp M_B$, (ii) the fermions remain gapped in the sub-manifolds at any value of $u$, and (iii) at any value of $u$, after the order parameter uniformly fluctuates in the sub-manifold, the protection symmetries in Table SM-I 
are restored. These sub-manifolds are shown schematically as the black circles in \Fig{combined}(b), where the direction of increasing $u$ is shown as the blue arrowed arc. Mathematically these sub-manifolds are given in the rows labeled as ``Sub-OPM'' in Table SM-IV 
of the supplementary material. The statement (iii) can be shown by the fact that when whole symmetries in Table SM-I 
act on any point in the sub-manifold it generates the entire sub-manifold.

The sub-manifolds in the preceding paragraph are in one-to-one correspondence with the fermion mass $M(u)$ in Table SM-IV
. 
The $Q^{\mathbb{R,C}}$ corresponding to $M(u)$ are given in the rows labeled by ``$M(u)\ra Q^{\mathbb{R,C}}$'' in Table SM-IV
. Note that the sub-manifolds formed by $Q^{\mathbb{R,C}}$ agree with the OPM in Table \ref{NLS2} of the main text.
Substituting these $Q^{\mathbb{R,C}}$ into $W_{\rm WZW}$ in Table SM-III 
computes the accumulated Berry's phase change as a function of $u$. It is straightforward to show that the result is $i \theta W$ where $W$ is given in Table \ref{NLS2} of the main text, and the value of $\theta$ depends on the final value of $u$. Tuning $u$ from $0$ to $1$ changes $\theta$ from $0$ to $2\pi$. The NL$\s$ with $\theta$ varying in $[0,2\pi]$ is the bosonic action when the fermion mass is tuned from $m_0>0$ to $m_0<0$. 

This derivation of $\theta$ term from the WZW term is analogous to that carried out in Ref.\cite{Abanov2000} to derive the O(3) NL$\s$ with $\theta=\pi$ term from the O(4) NL$\s$ model with WZW term in $D=1+1$.

{\bf{The symmetry protected critical spin liquids as the boundary of spin SPTs}}
In Ref.\cite{Huang2021} it is shown that the $d=1$ (OPM=$S^3$) and $d=2$ (OPM=$S^4$) NL$\s$ models with level-1 WZW term can be derived from the ``$\pi$-flux'' \cite{Affleck1988} massless fermionic ``spinon'' theory coupled to a ``charge''-$SU(2)$ gauge field, which model Mott insulators \cite{Lee2006}. The gamma matrices and the symmetries of these massless fermion theories are given in the $D=1+1$, and $D=2+1$ ``real class'' column\footnote{The reason that it is the real class rather than the complex class is due to the necessity of coupling spinons to the charge $SU(2)$ gauge field.} of Table SM-II
. The relevant flavor number is $n=4$ and $n=8$ for $D=1+1$ and $2+1$ respectively. The purpose of the charge SU(2) gauge field is to mediate confinement which enforces the no-double occupation constraint in Mott insulators so that the low energy degrees of freedom are spins rather than spinons. In the real class with $D=0+1$ and $n=4$, which was not discussed in Ref.\cite{Huang2021}, there are four Majorana fermion levels (which correspond to two, namely, spin up and spin down complex fermion levels). Under the charge-$SU(2)$ singlet constraint only one of the spin levels can be occupied. This gives rise to a spin 1/2. The coherent state path integral is the NL$\s$ model with OPM=$S^2$ and the WZW term. The gaplessness of the $d=0,1,2$ NL$\s$ models is protected by the $O(3)$, $O(4)$ 
and $O(5)$ symmetries. The OPM and WZW term of these NL$\s$ models are given in Table SM-V 
of the supplementary material. These critical spin liquids can be realized at the boundary of $d=1,2,3$ spin SPTs whose NL$\s$ models and the associated $\theta$ terms are given in Table SM-V 
of the supplemental material. 

{\bf{Conclusion}}
We have bosonized a class of $\mathbb{Z}$-classified fermion SPTs. This unifies the continuum field theory descriptions of bosonic and fermionic SPTs (the bosonization for $\mathbb{Z}_2$-classified SPTs will be left for a future publication). The results are NL$\s$ models with topological $\theta$ terms. 
In particular, the trivial SPT corresponds to $\theta=0$ and the non-trivial SPT corresponds to $\theta=2\pi$. Tuning the $\theta$ value from $0$ to $2\pi$ triggers the SPT phase transition. Finally, we present the spin NL$\s$ models whose boundaries are symmetry-protected critical spin liquids.

\bibliographystyle{ieeetr}
\bibliography{bibs}

\end{document}


\restylefloat{table}

\title{Supplemental materials: Non-abelian bosonization of fermion symmetry-protected topological state} 
\author{Yen-Ta Huang$^{1}$}
\author{Dung-Hai Lee$^{1,2}$}\email{Corresponding author: dunghai@berkeley.edu}

\affiliation{$^1$ Department of Physics, University of California, Berkeley, CA 94720, USA.\\
$^2$ Materials Sciences Division, Lawrence Berkeley National Laboratory, Berkeley, CA 94720, USA.
}
\maketitle

Throughout the supplemental material if we refer to a table or an equation in the main text it will be explicitly mentioned. Reference to tables/equations in the same supplemental material will not be explicitly mentioned. In the following $I,X,Y,Z$ stands for Pauli matrices $\s_0,\s_x,\s_y,\s_z$. When one Pauli matrix stands next to another one it indicates a tensor product.

\section{Protection symmetry generators, $\G$ matrices, and the mass term of the fermion SPTs bosonized in Table II 
of the main text}
In this section, we summarize the fermion SPTs that we bosonize. These SPTs are all $\mathbb{Z}$ classified as shown in Ref. \cite{Huang2021}.
\begin{table}[H]
\centering
\small
\begin{tabular}{ |c|c|c| }
\hline
$D=1+1$ & Real class & Complex class \\
\hline
$\Gamma_1$ & $X\otimes I_n $ & $X\otimes I_n$ \\
\hline
Protection symmetry generators 
& \thead{$T=Z\otimes I_n$\\ $O(n) : I\otimes g$ \\ 
where $g \in O(n)$} 
& \thead{ $T = Z\otimes I_n$ \\ $C=I\otimes I_n$\\ 
$U(n) : I \otimes g$ \\ 
where $g \in U(n)$} \\
\hline
${\rm Mass}~M_B$ & $Y\otimes I_n$ & $Y\otimes I_n$ \\
\Xhline{3\arrayrulewidth}
$D=2+1$ & Real class & Complex class \\
\hline
$\Gamma_1,\G_2$ & $ZZ\otimes I_n, XI\otimes I_n $ & $ZZ\otimes I_n, XI\otimes I_n $ \\
\hline
Protection symmetry generators 
& \thead{$T = ZX \otimes I_n$ \\ 
$O_+(n) \times O_-(n): IP_+ \otimes g_+ + IP_- \otimes g_-$ \\ 
where $g_+ \in O_+(n)$ and $g_- \in O_-(n)$} 
& \thead{ $T = ZX \otimes I_n$ \\ $C=IZ \otimes I_n$ \\ 
$U_+(n) \times U_-(n): IP_+ \otimes g_+ + IP_- \otimes g_-$ \\ 
where $g_+ \in U_+(n)$ and $g_- \in U_-(n)$} \\
\hline
${\rm Mass~}M_B$ & $YI\otimes I_n$ & $YI\otimes I_n$ \\
\Xhline{3\arrayrulewidth}
$D=3+1$ & Real class & Complex class \\
\hline
$\Gamma_1,\G_2,\G_3$ & $ZZ \otimes I_n, ZX \otimes I_n, XI\otimes I_n$ & $ZZ \otimes I_n, ZX \otimes I_n, XI\otimes I_n$ \\
\hline
Protection symmetry generators 
& \thead{$T =ZE \otimes I_n$ \\ 
$O(n) : II \otimes g$ \\ 
where $g \in O(n)$} 
& \thead{$T = ZE \otimes I_n$ \\ 
$C=II\otimes I_n$ \\ 
$U(n) : II \otimes g$ \\ 
where $g \in U(n)$} \\
\hline
${\rm Mass}~M_B$ & $YI\otimes I_n$ & $YI\otimes I_n$ \\
\hline
\end{tabular}
\caption{Protection symmetry generators, $\G$ matrices, and the mass term of fermions SPT in $(1+1)$-D, $(2+1)$-D, and $(3+1)$-D. $P_\pm:=(I\pm Z)/2$. These fermion SPTs are bosonized in Table II 
	of the main text.} 
\label{tab:sym}
\end{table}

\section{The gamma matrices ($\g_i$), and the protection symmetry generators at the boundary of the SPT in Table \ref{tab:sym}
}
Table \ref{tab:dw} summarizes the gamma matrices and the symmetry generators of the gapless fermions living on the boundary of the SPTs in Table \ref{tab:sym}. The Hamiltonian for these gapless fermions is given by Eq. 2 
of the main text. 
\begin{table}[H]
\centering
\small
\begin{tabular}{ |c|c|c| }
\hline
$D=0+1$ & Real class & Complex class \\
\hline
Boundary $\g$ matrices & $\sim$ & $\sim$ \\
\hline
Boundary symmetry generators & \thead{$T=1$\\ $O(n) : g$ \\ 
where $g \in O(n)$} 
& \thead{ 
$T=1$\\ $C=1$\\ $U(n) : g$ \\ 
where $g \in U(n)$} \\
\Xhline{3\arrayrulewidth}
{$D=1+1$} & Real class & Complex class \\
\hline
Boundary $\g$ matrices $\g_1$ & $Z\otimes I_n$ & $Z\otimes I_n$ \\
\hline
Boundary symmetry generators &\thead{$T = X \otimes I_n$ \\ 
$O_+(n) \times O_-(n): P_+ \otimes g_+ + P_- \otimes g_-$ \\ 
where $g_+ \in O_+(n)$ and $g_- \in O_-(n)$} 
& \thead{ $T = X \otimes I_n$ \\ $C=Z \otimes I_n$ \\ 
$U_+(n) \times U_-(n): P_+ \otimes g_+ + P_- \otimes g_-$ \\ 
where $g_+ \in U_+(n)$ and $g_- \in U_-(n)$} \\
\Xhline{3\arrayrulewidth}
$D=2+1$ & Real class & Complex class \\
\hline
Boundary $\g$ matrices $\g_1,\g_2$ & $Z \otimes I_n, X \otimes I_n$ & $Z \otimes I_n, X \otimes I_n$ \\
\hline
Boundary symmetry generators &\thead{$T =E \otimes I_n$ \\ 
$O(n) : I \otimes g$ \\ 
where $g \in O(n)$} 
& \thead{$T = E \otimes I_n$ \\ 
$C=I\otimes I_n$ \\ 
$U(n) : I \otimes g$ \\ 
where $g \in U(n)$} \\
\hline
\end{tabular}
\caption{The boundary gamma matrices $\g_i$, and the boundary protection symmetry generators, for the massless fermion theories on the boundary of the SPTs in Table \ref{tab:sym}.} 
\label{tab:dw}
\end{table}
\section{Bosonized NL$\s$ models for the critical theories describing the phase transition between trivial and non-trivial SPTs in Table \ref{tab:sym}}
The massless fermion Hamiltonian that are bosonized in the following table corresponds to $m_0=0$ in Eq. 1 
of the main text. These fermion theories describe the critical point of the $m_0>0$ to $m_0<0$ SPT phase transitions in Table \ref{tab:sym}.

\begin{table}[H]
\centering
\small
\renewcommand{\arraystretch}{2}
\begin{tabular}{ |c|c|c| }
\hline
$D=1+1$ & Real class & Complex class \\
\hline
Bosonized critical theory $S=$ &$\frac{1}{16\pi} \int_{\mathcal{M}} d^2x \, \tr\left[ \partial_\mu Q^{\mathbb{R}} \partial^\mu (Q^{\mathbb{R}})^T \right] -W_{\rm WZW}$&
$
\frac{1}{8\pi} \int_{\mathcal{M}} d^2x \, \tr\left[ \partial_\mu Q^{\mathbb{C}} \partial^\mu Q^{\mathbb{C}\dagger} \right]-W_{\rm WZW}$\\
\hline
$W_{\rm WZW}=$ &\thead{$\frac{2 \pi i}{48\pi^2} \int\limits_{\mathcal{B}} \, \tr \Big\{ \left[(\tilde{Q}^{\mathbb{R}})^T d \tilde{Q}^{\mathbb{R}}\right]^3 \Big\}$} &$\frac{2\pi i}{24\pi^2} \int\limits_{\mathcal{B}} \, \tr \Big[ \left( \tilde{Q}^{\mathbb{C}\dagger} d \tilde{Q}^{\mathbb{C}} \right)^3 \Big]$\\
\hline
OPM&$Q^{\mathbb{R}}\in O(n)$ &$Q^{\mathbb{C}}\in U(n)$ \\
\Xhline{3\arrayrulewidth}
$D=2+1$ & Real class & Complex class \\
\hline
Bosonized critical theory model $S=$ &${1\over 2g} \int_{\mathcal{M}} d^3 x \, \tr\left[ \partial_{\mu} Q^{\mathbb{R}} \partial^{\mu}Q^{\mathbb{R}} \right]-W_{\rm WZW}$&
$
{1\over 2g} \int_{\mathcal{M}} d^3 x \, \tr\left[ \partial_{\mu} Q^{\mathbb{C}} \partial^{\mu}Q^{\mathbb{C}} \right]-W_{\rm WZW}$\\
\hline
$W_{\rm WZW}=$ &$\frac{2 \pi i}{512 \pi^2} \int\limits_{\mathcal{B}} \tr \left[ \tilde{Q}^{\mathbb{R}} \, (d \tilde{Q}^{\mathbb{R}})^4 \right]$ &$\frac{ 2 \pi i }{256 \pi^2} \int\limits_{\mathcal{B}} \, \tr \left[\tilde{Q}^{\mathbb{C}} \,\left( d \tilde{Q}^{\mathbb{C}} \right)^4 \right]$
\\
\hline
OPM &$Q^{\mathbb{R}}\in \frac{O(n)}{O(n/2) \times O(n/2)}$&$Q^{\mathbb{C}}\in \frac{U(n)}{U(n/2)\times U(n/2)}$ \\
\Xhline{3\arrayrulewidth}
$D=3+1$ & Real class & Complex class \\
\hline
Bosonized critical theory $S=$ &${1\over 2g} \int_{\mathcal{M}} d^3 x \, \tr\left[ \partial_{\mu} Q^{\mathbb{R}} \partial^{\mu}Q^{\mathbb{R}} \right]-W_{\rm WZW}$&
$
{1\over 2g} \int_{\mathcal{M}} d^3 x \, \tr\left[ \partial_{\mu} Q^{\mathbb{C}} \partial^{\mu}Q^{\mathbb{C}} \right]-W_{\rm WZW}$\\
\hline
$W_{\rm WZW}=$ &$\frac{2\pi}{960\pi^3} \int\limits_{\mathcal{B}} \, \tr \Big[ \left(\tilde{Q}^{\mathbb{R}\dagger} d \tilde{Q}^{\mathbb{R}} \right)^5 \Big]$ &$\frac{2\pi}{480\pi^3} \int\limits_{\mathcal{B}} \, \tr \Big[\left(\tilde{Q}^{\mathbb{C}\dagger} d\tilde{Q}^{\mathbb{C}} \right)^5 \Big]$
\\
\hline
OPM &$Q^{\mathbb{R}}\in {U(n)\over O(n)}$&$Q^{\mathbb{C}}\in U(n)$\\
\hline
\end{tabular}
\caption{A summary of the bosonized theory for the $M_B=0$ in Table \ref{tab:sym} of the supplemental material. These are the critical theories for SPT transitions.} 
\label{tab:crt}\end{table}

\section{The one parameter family of sub-manifolds correspond to $M(u)$ }
The following table contains the information needed to derive the NL$\s$ with $\theta$ term from the critical theory in Table \ref{tab:crt}.
\begin{table}[H]
\centering
\small
\renewcommand{\arraystretch}{2}
\begin{tabular}{ |c|c|c| }
\hline
$D=1+1$ & Real class & Complex class \\
\hline
$M(u)$&\thead{$\sin \left({\pi\over 2}f(u)\right)M_B+\cos \left({\pi\over 2}f(u)\right)ZH_A(\t,x)$}&\thead{$\sin \left({\pi\over 2}f(u)\right)M_B+\cos \left({\pi\over 2}f(u)\right)ZH(\t,x)$} \\
\hline
$M(u)\ra Q^{\mathbb{R,C}}$ & \thead{$\tilde{Q}^{\mathbb{R}}=\sin \left({\pi\over 2}f(u)\right)I_n+i\cos \left({\pi\over 2}f(u)\right)H_A(\t,x)$}& \thead{$\tilde{Q}^{\mathbb{C}}=\sin \left({\pi\over 2}f(u)\right)I_n+i\cos \left({\pi\over 2}f(u)\right)H(\t,x)$} \\
\hline
Sub-OPM&$H_A\in {O(n)\over U(n/2)}$&$H\in {U(n)\over U(n/2)\times U(n/2)}$\\
\Xhline{3\arrayrulewidth}
$D=2+1$ & Real class & Complex class \\
\hline
$M(u)$&\thead{$\sin \left({\pi\over 2}f(u)\right)M_B+\cos \left({\pi\over 2}f(u)\right)Y\Big[XS(\t,x,y)-iYA(\t, x,y)\Big]$}&
\thead{$\sin \left({\pi\over 2}f(u)\right)M_B+\cos \left({\pi\over 2}f(u)\right)Y\Big[XH_1(\t,x,y)+YH_2(\t, x,y)\Big]$} \\ 
\hline
$M(u)\ra Q^{\mathbb{R,C}}$ & \thead{$\tilde{Q}^{\mathbb{R}}=\sin \left({\pi\over 2}f(u)\right)ZI_{n}+\cos \left({\pi\over 2}f(u)\right)\Big[XS(\t,x,y)$\\$-iYA(\t, x,y)\Big]$}& \thead{$\tilde{Q}^{\mathbb{C}}=\sin \left({\pi\over 2}f(u)\right)ZI_{n}+\cos \left({\pi\over 2}f(u)\right)\Big[XH_1(\t,x,y)$\\$+YH_2(\t, x,y)\Big]$}\\
\hline
Sub-OPM&$S+A\in O(n)$&$H_1+iH_2\in U(n)$\\
\Xhline{3\arrayrulewidth}
$D=3+1$ & Real class & Complex class \\
\hline
$M(u)$&\thead{$\sin \left({\pi\over 2}f(u)\right)M_B+\cos \left({\pi\over 2}f(u)\right)YXS(\t,x,y,z)$}&\thead{$\sin \left({\pi\over 2}f(u)\right)M_B+\cos \left({\pi\over 2}f(u)\right)YXH(\t,x,y,z)$} \\ 
\hline
$M(u)\ra Q^{\mathbb{R,C}}$ & $\tilde{Q}^{\mathbb{R}}=\sin \left({\pi\over 2}f(u)\right)I_n+\cos \left({\pi\over 2}f(u)\right)S(\t,x,y,z)$& $\tilde{Q}^{\mathbb{C}}=\sin \left({\pi\over 2}f(u)\right)I_n+\cos \left({\pi\over 2}f(u)\right)H(\t,x,y,z)$\\
\hline
Sub-OPM&$S\in {O(n)\over O(n/2)\times O(n/2)}$&$H\in {U(n)\over U(n/2)\times U(n/2)}$\\
\hline
\end{tabular}
\caption{The mass terms $M(u)$ is chosen so that $u=0,1$ realizes $\mp M_B$ in Table \ref{tab:sym}. The rows labeled by ``$M(u)\ra Q^{\mathbb{R,C}}$'' give the mapping from $M(u)$ to the OPM in Table \ref{tab:crt}. When the order parameter fluctuates uniformly in the sub-manifolds given in the rows labeled by ``Sub-OPM'', the protection symmetries in Table \ref{tab:sym} are restored.} 
\label{tab:mu}\end{table}

\section{The NL$\s$ models for the critical spin liquids and their bulk spin SPTs} 
In the following table, the NL$\s$ models describing the critical spin liquids and their corresponding bulk spin SPTs are given.

\begin{table}[H]
\centering
\small
\renewcommand{\arraystretch}{2}
\begin{tabular}{|c|c|c|}
\hline
~&Boundary $D=0+1$ &Bulk $D=1+1$\\
\hline
NL$\s$ model $S$ &$ \frac{1}{2g} \int\limits_{\mathcal{M}_{0+1}} d\t \, \left( \partial_\mu \hat{\Omega} \right)^2-W_{\rm WZW}$&$ \frac{1}{2g} \int\limits_{\mathcal{M}_{1+1}} d^2x \, \left( \partial_\mu \hat{\Omega} \right)^2-W_\theta$\\
\hline
$W_{\rm WZW}$ or $W_\theta$ & $W_{\rm WZW}=$$\frac{2\pi i }{8 \pi} \int\limits_{\mathcal{B}} \epsilon^{ijk} \tilde{\Omega}_i \, d\tilde{\Omega}_j \, d\tilde{\Omega}_k $&$W_\theta=\frac{2\pi i}{8 \pi} \int\limits_{\mathcal{M}_{1+1}} \epsilon^{ijk} \Omega_i \, d\Omega_j \, d\Omega_k$ \\
\hline
OPM &$\hat{\Omega}\in S^2$&$\hat{\Omega}\in S^2$\\
\Xhline{3\arrayrulewidth}
~&Boundary $D=1+1$ &Bulk $D=2+1$\\
\hline
NL$\s$ model $S$ &$ \frac{1}{4\pi} \int\limits_{\mathcal{M}_{1+1}} d^2 x \, \left( \partial_\mu \hat{\Omega} \right)^2-W_{\rm WZW}$&$ \frac{1}{2g} \int\limits_{\mathcal{M}_{2+1}} d^3x \, \left( \partial_\mu \hat{\Omega} \right)^2-W_\theta$\\
\hline
$W_{\rm WZW}$ or $W_\theta$& $W_{\rm WZW}=$ $\frac{2\pi i }{12 \pi^2} \int\limits_{\mathcal{B}} \epsilon^{ijkl} \tilde{\Omega}_i \, d\tilde{\Omega}_j \, d\tilde{\Omega}_k \, d\tilde{\Omega}_l$&$W_\theta=\frac{2\pi i}{12 \pi^2} \int\limits_{\mathcal{M}_{2+1}} \epsilon^{ijkl} \Omega_i \, d\Omega_j \, d\Omega_k \, d\Omega_l$ \\
\hline
OPM &$\hat{\Omega}\in S^3$&$\hat{\Omega}\in S^3$\\
\Xhline{3\arrayrulewidth}
~&Boundary $D=2+1$ &Bulk $D=3+1$\\
\hline
NL$\s$ model $S$ &$\frac{1}{2g} \int\limits_{\mathcal{M}_{2+1}} d^3 x \, \left( \partial_\mu \hat{\Omega} \right)^2-W_{\rm WZW}$&$ \frac{1}{2g} \int\limits_{\mathcal{M}_{3+1}} d^4x \, \left( \partial_\mu \hat{\Omega} \right)^2-W_\theta$\\
\hline
$W_{\rm WZW}$ or $W_\theta$& $W_{\rm WZW}=$ $\frac{2\pi i }{64 \pi^2} \int\limits_{\mathcal{B}} \epsilon^{ijklm} \tilde{\Omega}_i \, d\tilde{\Omega}_j \, d\tilde{\Omega}_k \, d\tilde{\Omega}_l \, d\tilde{\Omega}_m.$&$W_\theta=\frac{2\pi i}{64 \pi^2} \int\limits_{\mathcal{M}_{3+1}} \epsilon^{ijklm} \Omega_i \, d\Omega_j \, d\Omega_k \, d\Omega_l \, d\Omega_m.$ \\
\hline
OPM &$\hat{\Omega}\in S^4$&$\hat{\Omega}\in S^4$\\
\hline
\end{tabular}
\caption{The NL$\s$ models for the critical spin liquids and their bulk spin SPTs. } 
\label{tab:Mott}
\end{table}

\bibliographystyle{ieeetr}
\bibliography{bibs}